%% file: main.tex
\documentclass[sigconf]{acmart}

\usepackage{cleveref}
\usepackage{ifthen}
\usepackage{subcaption}
\usepackage{booktabs}   
\usepackage{xspace}
\usepackage{tabularx}
\usepackage{colortbl}
\usepackage{tcolorbox}
\usepackage{soul}

\AtBeginDocument{%
  }

\copyrightyear{2027}
\acmYear{2027}
\setcopyright{cc}
\setcctype{by}
\acmConference[SIGCSE TS 2027] {Proceedings of the 58th ACM Technical Symposium on Computer Science Education V.1}{February 17--20, 2027}{Sacramento, CA, USA.}
\acmBooktitle{Proceedings of the 58th ACM Technical Symposium on Computer Science Education V.1 (SIGCSE TS 2027), February 17--20, 2027, Sacramento, CA, USA}
\acmISBN{979-8-4007-3075-7/2027/02}
\acmDOI{10.1145/3857222.3860139}

\begin{document}

\title[Understanding Computing Instructors’ Perceptions of Student Sense of Belonging]{Understanding Computing Instructors’ \\Perceptions of Student Sense of Belonging}


\author{Ritesh Kanchi}
\email{rkanchi@.g.harvard.edu}
\orcid{0009-0006-7978-0821}
\affiliation{
    \institution{Harvard University}
    \city{Cambridge}
    \state{Massachusetts}
    \country{USA}
}
\authornote{The first two authors are equal contributors to this work.}

\author{Teresa Luo}
\email{teresaluo@g.harvard.edu}
\orcid{0009-0009-9099-7767}
\affiliation{
    \institution{Harvard University}
    \city{Cambridge}
    \state{Massachusetts}
    \country{USA}
}
\authornotemark[1]

\author{Sam Grunebaum}
\email{samuel_grunebaum@mde.harvard.edu}
\orcid{0009-0003-6457-5970}
\affiliation{
    \institution{Harvard University}
    \city{Cambridge}
    \state{Massachusetts}
    \country{USA}
}

\author{Alex Oesterling}
\email{aoesterling@g.harvard.edu}
\orcid{0000-0001-8546-0089}
\affiliation{
    \institution{Harvard University}
    \city{Cambridge}
    \state{Massachusetts}
    \country{USA}
}

\newif\ifcomments
\commentstrue   

\newcommand{\alex}[1]{{\ifcomments\textcolor{orange}{[AO: #1]}\normalfont\fi\xspace}}
\newcommand{\rk}[1]{{\ifcomments\textcolor{blue}{[RK: #1]}\normalfont\fi\xspace}}
\newcommand{\sg}[1]{{\ifcomments\textcolor{green}{[SG: #1]}\normalfont\fi\xspace}}
\newcommand{\tl}[1]{{\ifcomments\textcolor{red}{[TL: #1]}\normalfont\fi\xspace}}
\newcommand{\kg}[1]{{\ifcomments\textcolor{violet}{[KG: #1]}\normalfont\fi\xspace}}
\newcommand{\pnine}[1]{{#1}}
\newcommand{\imp}[1]{{\hl{#1}\xspace}}

\newcommand{\sob}{sense of belonging}

\newcommand{\etal}{\textit{et al. }}
\newcommand{\ie}{\textit{i.e.,}}
\newcommand{\eg}{\textit{e.g.,}}

\begin{abstract}
  \input{00-abstract}
\end{abstract}

\ccsdesc[500]{Social and professional topics~Computing education}

\keywords{ Belonging, Introductory Computing, Computer Science Education}


\maketitle

\input{01-intro}

\input{02-related}

\input{03-methodology}
\input{04-new-findings}

\input{05-new-discussion}
\input{06-conclusion}

\begin{acks}
We thank Krzysztof Gajos, Hongjin Lin, Jianna So, Lauren Bricker, Kevin Lin, and the Spring 2026 class of CS2760, whose feedback and support shaped our research and writing.
\end{acks}

\bibliographystyle{ACM-Reference-Format}
\bibliography{bibliography}

\appendix

\end{document}

%% file: 00-abstract.tex
Students' sense of belonging has been described as a feeling of community acceptance, respect, inclusion, and membership. In computing, student belonging is shaped by psychological, social, academic, and environmental factors, including their relationships with instructors. However, despite prior work emphasizing instructors' influence on student belonging, limited research has examined how computing instructors perceive, understand, and attempt to support belonging. Through nine semi-structured qualitative interviews with introductory computing instructors at large public universities in the United States, we investigated how instructors define, measure, and conceptualize belonging; what interventions they use to support student belonging; and what barriers limit the efficacy of these interventions. Our findings show that instructors defined belonging through value and identity representation, self-efficacy, and community; relied on surveys, observations, and student performance to measure belonging; saw belonging to be influenced by prior experience, major, student and institutional culture, and representation; implemented student-oriented interventions and designed courses around belonging; and identified multi-level challenges to supporting belonging. These results highlight opportunities to design belonging interventions that attend to both prior work and instructor conceptualizations, while also considering tensions in utilizing teaching assistants, balancing students' expectations, and adapting to broader societal trends affecting belonging.

%% file: 01-intro.tex
\section{Introduction}
Sense of belonging has been broadly understood as students' subjective sense of fit, acceptance, value, and inclusion within the computing community~\cite{mahar2013conceptualizing, perlmutter-2023-field-accepted}. Computing education researchers have extended this understanding across multiple levels of analysis, such as students' belonging within a course, department, institution, and the field of computing as a whole~\cite{sax-2018-belonging-intro-women, moudalya-2021-measuring-students, runa2025sense}. Belonging is especially important for developing students' interest in pursuing computer science (CS) and persisting in the field~\cite{moudalya-2021-measuring-students, lewis-alignment-2019, master-she-belongs-2016}. Prior research has surfaced critical insights on what factors impact belonging, how belonging is experienced inequitably, and what interventions foster greater \sob{} in computing. While widely studied, there is limited research on computing instructors' perceptions of how belonging is manifested and influenced in their courses. This is an important gap as instructors hold both the intention and responsibility to translate broad belonging goals into concrete course practices. Such work is especially challenging in large introductory computing courses, where instructors are often limited by scale and resources~\cite{luo-student-support-2026, perlmutter-2023-field-accepted, alvarado2017microclasses}. In this paper, we bridge prior theoretical understanding of student belonging with experiences and insights from practitioners. We ask the following research questions: 

\begin{itemize}
    \item[\textbf{RQ1}] How do computing instructors conceptualize and assess students’ sense of belonging in their courses?
    \item[\textbf{RQ2}] How do instructors support belonging? 
    \item[\textbf{RQ3}] What constraints shape instructors' ability to support belonging at scale?
\end{itemize}

To answer these questions, we conducted nine semi-structured qualitative interviews with introductory computing instructors at four large R1 public universities in the United States. From our findings, we examine alignment and differences between instructor and prior work's conceptualizations of belonging to better inform intervention design. We also discuss opportunities to mobilize teaching assistants (TAs) in belonging at scale, account for students' expectations of computing, and respond to evolving societal trends affecting belonging in computing.

%% file: 02-related.tex
\section{Related Work}
\subsection{Student-Centered Belonging in CS}

Prior work on \sob{} in computing has predominantly centered students' reported experiences of belonging, often through student-reported surveys, including course-level belonging scales and multidimensional measurements adapted from STEM education~\cite{sax-2018-belonging-intro-women, lewis-alignment-2019, runa-2023-student-gender, moudalya-2021-measuring-students}. This work has been critical for understanding the psychological, social, and environmental factors that shape belonging in computing~\cite{runa2025sense}, including students' perception of identity and characteristic stereotypes about computing~\cite{lewis2016dontcode}, representation of identity and value in faculty and peers~\cite{lewis-alignment-2019, vakil2020scared, runa-2023-student-gender}, self-efficacy~\cite{veilleux2013relationship, blaney2017examining}, institutional structure~\cite{nguyen-2020-competitive, runa2025sense}, and social relationships at the course, department, and institution level~\cite{runa2025sense}.  

Recent work has increasingly highlighted that students minoritized by gender, race, sexuality, religion, and limited prior computing education face persistent inequities in belonging~\cite{runa-2023-student-gender, nguyen-2020-competitive, krause-levy-2021-cs1-belonging, lunn-2026-occupation, allen2018schools, lewis-alignment-2019, master-she-belongs-2016}. Critical and intersectional perspectives further suggest that these inequities are produced through the norms and power structures of computing education itself, including exclusionary classroom climates~\cite{barker2002defensive}, cultural stereotype framing computing ability as innate~\cite{master-she-belongs-2016}, epistemic oppression~\cite{rankin2021blacklong}, and misalignment between students' personal values and the dominant values in computing~\cite{vakil2020scared, agyare2026some}. This line of work particularly highlights how the broader institutions within computing can differentially legitimize particular identities and values. Consequently, computing education researchers have advocated for intersectional and context-dependent interventions to improve belonging at a broad, institutional level~\cite{runa2025sense}, where efforts to broaden participation should not only consider social barriers to enter computing but also to sustain agency and aspirations outside of computing~\cite{agyare2026some}.

Instructors influence how cultural norms, power dynamics, and institutional structures of computing education are interpreted and reflected within courses. Since instructors are central to supporting belonging in practice, we extend this prior work by examining how they define, assess, and conceptualize \sob{} (RQ1).

\subsection{Instructional Influences on Belonging}



Instructors' and course staff's values and behavior have been shown to shape student belonging~\cite{sax-2018-belonging-intro-women, runa2025sense, perlmutter-2023-field-accepted}. When instructors appear to endorse a fixed mindset about students’ capabilities, students experience a lower sense of belonging~\cite{canning2022professors}. In addition, prior work highlights the critical role instructors play in addressing intersecting oppression in computing education, arguing that policy changes alone are inadequate if instructors' own biases and prejudices continue to shape students' classroom experiences~\cite{rankin2021blacklong}. In contrast, students' perceptions of personal instructor connections (\eg{} instructors' knowing students' names, instructor immediacy practices) have been shown to increase belonging~\cite{romney2025student, cowit2023teaching}. Prior work has also identified specific instructional practices that support belonging in computing, including more frequent instructor-student interactions~\cite{blaney2017examining}, inclusive language~\cite{sax-2018-belonging-intro-women}, collaborative pedagogies~\cite{fong-exploring-2024}, and culturally responsive teaching (CRT) practices~\cite{vonbriesen2025interventions}. 

In large computing courses, students may also encounter instructional values indirectly through TAs. Prior work has shown that TA-student interactions substantially influence students' \sob{} through peer support and shaping of course culture~\cite{alvarado2017microclasses, perlmutter-2023-field-accepted, brooksramirez2026pluriversal, kanchi-systems-2026}. Recent work on TA training rooted in inclusive, accessible, and culturally responsive and relevant pedagogies~\cite{yan2025teaching, kanchi-systems-2026} further advocates for the importance of preparing course staff for a reflective teaching culture to cultivate belonging in computing.

Instructional practices reflect both the broader cultural and institutional values that shape how students understand their place in computing. Therefore, understanding instructors' values and actions, alongside course staff who may be perceived as extensions of said values, is critical for understanding how \sob{} can be cultivated. We examine this by identifying the belonging-rooted values instructors prioritize in their teaching practices and how these values are reflected in their efforts to support belonging (RQ2). We also contextualize these practices within the challenges and constraints instructors face in implementing them (RQ3).

%% file: 03-methodology.tex
\section{Methods}

\newcolumntype{Y}{>{\raggedright\arraybackslash}X}
\begin{table*}[t]\centering
\centering
\caption{Interview Participant Demographics.} 
\Description{This is a table of Interview Participant demographics. This table has 10 rows and 6 columns. The first row is a header row with the following columns: ID, Gender, Courses taught, Students taught, Years taught, and School Location.}
\small
\rowcolors{2}{gray!10}{white}
\begin{tabularx}{\linewidth}{l l l l l X}
\toprule
\textbf{ID} & \textbf{Gender} & \textbf{Courses taught} & \textbf{Students taught} & \textbf{Years taught }& \textbf{Institution}\\
\midrule
P1 & F  & CS1, CS2, Discrete Math & Majors/Non-Majors & 4 & West-Coast University 1 (WCU1) \\
P2 & M  & Data Structures \& Algorithms, Data Programming & Non-Majors  & 7 & West-Coast University 1 (WCU1) \\
P3 & M  & CS0, CS1, CS2, Computer Architecture  & Majors/Non-Majors & 1 & West-Coast University 2 (WCU2) \\
P4 & M  & CS1, CS2 &  Majors/Non-Majors & 3 & West-Coast University 1 (WCU1) \\
P5 &  M  & Data Structures, Discrete Math & Majors/Non-Majors & 15 & West-Coast University 2 (WCU2) \\
P6 &  F & CS2 (Non-Majors), Computer Architecture (Majors) & Majors/Non-Majors  & 1 & Southern University (SU) \\
P7 &  F  & CS1 & Non-Majors & 14 & Midwestern University (MWU) \\
P8 &  M & CS1, CS2 &  Majors/Non-Majors & 9 & West-Coast University 1 (WCU1)\\
P9 &  M & CS0, Computer Architecture (Majors) & Majors/Non-Majors & 25 & West-Coast University 2 (WCU2)\\
\bottomrule
\end{tabularx}
\label{tab:interview-demographics}
\end{table*}



We conducted an interview study comprising of nine (three female, six male) introductory computing instructors from large public U.S. universities (Table~\ref{tab:interview-demographics}). Each semi-structured remote interview lasted approximately 30 minutes per participant. Eight participants held teaching-focused faculty appointments, while P3 was a lecturer. Participants taught large lecture-based courses supported by TAs who facilitated smaller sections or labs. Using convenience sampling, we contacted instructors teaching at undergraduate institutions attended by research-team members. To avoid potential familiarity bias, team members interviewed and coded transcripts of participants with no prior relation. The study was reviewed and approved by the Harvard University Committee on the Use of Human Subjects (Protocol IRB26-0414).

Our research questions guided our interview questions for all participants, covering (1) their courses; (2) their understanding of \sob{}, measurement strategies, and potential interventions; and (3) speculative \sob{} improvements, institutional factors and constraints, and department-level discussions. 
We analyzed transcripts using inductive thematic analysis~\cite{braun_using_2006}. 
The interviewer and a second researcher who had not conducted that interview independently coded each transcript, applying codes from earlier transcripts where they fit and creating new codes where they did not, so the codebook evolved throughout coding. After each transcript, the two coders met to discuss their codes and interpretations. 
After all transcripts were coded, the full research team then reviewed the codebook together, merged overlapping codes, and used affinity diagramming to organize codes into six themes that structured our findings and addressed different RQs.

%% file: 04-new-findings.tex
\section{Findings}
\label{sec:findings}

\subsection{Defining Students' Sense of Belonging (RQ1)}
\subsubsection{Value and identity representations} Several participants defined \sob{} as how students' personal values, identities, and experiences were represented: ``Can [students] come in with their own backgrounds, their own interests, their own experiences, their own perceptions?'' (P8) Other definitions connected belonging to CRT pedagogy which considers students' intersectional identities when entering a course (P3), as well as the importance of considering how belonging may differ across minority and majority groups (P6). P9 illustrated this connection through his own identity, describing how being both a computer scientist and a Latino man shaped his conception of belonging---highlighting the importance of students seeing their identities represented in computing.


\subsubsection{Self-efficacy} Other participants defined \sob{} as ``[students having the] confidence that they can or should be able to be successful in [...] computer science.'' (P1) Such confidence is fostered through cultivating student trust in their ability to obtain technical competency (P5), trust in being able to learn from failures (P9), and trust in the course to help them succeed (P7).

\subsubsection{Community}
Lastly, some participants defined \sob{} as when ``students feel like they're in a safe and supportive environment to learn and [feel] like [community members] support them.'' (P4) P2 connected belonging with the theory of communities of practice~\cite{wenger1998communities}, which emphasizes the importance of social interactions in promoting positive learning experiences. 

\subsection{Measuring Student Belonging (RQ1)}
\label{sec:measuring}
\subsubsection{Surveys \& Reflections}
P3 described surveying students throughout the course (\eg{} weekly, milestone-based) as ``a doorway to [...] engage in communication and dialogue,'' especially when students may ``fall through the cracks'' in large courses. While a majority of participants did not explicitly survey belonging, they used reflexive proxy questions such as ``Do you see yourself as potentially [being a] computer scientist?'' (P5) and ``What is one thing that went poorly in the course for you this week?'' (P3). Several participants administered post-assignment reflections; however, since they could not read every reflection, they used TAs to surface indicative reflections (P1, P4): ``if a student says, `Hey, I'm really struggling', or `I spent 20 hours on this assignment', or `I'm about to crash out' [...] a TA will flag that, and then I will often follow up with the student.'' (P4) However, participants felt that open-ended surveys and reflections may not surface ``honest'' and ``vulnerable'' data (P1, P6). Some participants voluntarily brought in institutional third-parties to conduct course-wide surveys for greater transparency (P4, P8). 




\subsubsection{Observation} 
Participants directly and indirectly observed students who needed support. In some instances, this was noticing when students who often attended office hours or lecture disappeared suddenly (P1)---prompting an immediate check-in (P1, P8). Participants also discussed how they flagged academic performance (\eg{} poor grades, late assignments, extensions) and attendance (notably, not participation) to check in with students (P3, P7): ``I reach out to students who aren't [...] coming to early classes, [or] completing the mastery-based homework that I want them to get 100\% on.'' (P7) P9 used grades to find students who ``need a little bit of support'' and should be recommended to take an earlier introductory course first. However, P1 observed a gap in knowing how many students ``don't feel like they can even take the course.'' Since participants could not interface with every student, TAs provided closer observation (P1, P3, P4, P5). During course staff meetings, P3 asked questions such as ``[how] are the students doing?'' and ``do they feel like they belong?'' P5 described a TA-led ``student support team'' who identified struggling students and ``[offered] sit-down sessions where [students] can meet with staff.''


\subsection{Identifying Influences on Belonging (RQ1)}
\label{sec:factors}

\subsubsection{Prior Computing Experience}
\label{sec:prior-experience}
Introductory courses are often where students new to computing begin with limited technical knowledge and confidence~\cite{sax-2018-belonging-intro-women}. P1 noted, ``I really expect that [a lower \sob{}] could be even more prevalent'' for novices, further referencing her own student experience: ``I definitely had that perception of, `I'm [the] dumbest person in the room, everyone's gonna be so much smarter than me and have so much more experience.''' Students' diverse interests, goals, and prior experience were said to make calibrating content difficult (P1, P3, P5) (further discussed in Section~\ref{sec:relevance}). P9 also identified a correlation between prior experience and grades, noting that his students have said ``I got a C or a B [...] I guess computing isn't for me.''

 

 


\subsubsection{Major} 
\label{sec:major}
For CS majors, participants described belonging as \textit{feeling} \textit{like} a CS student, with their department being their home. P7 emphasized this importance: ``If I was teaching a majors class, I really want to have them feel a \sob{} in computing.'' In contrast, when teaching non-majors, P7 noted that since the CS department ``is not their primary home,'' there is a greater need to support non-majors' ``self-efficacy in [the] class.'' However, for non-majors who develop an interest in CS through their introductory courses, there can be limited pathways to transfer or even clear expectations that they \textit{cannot} study CS (P2, P4, P6), leading them to feel excluded before further pursuing CS.




\subsubsection{Student Culture} \label{sec:student-culture}
Student culture defines the expectations about who belongs in computing and how students should approach courses.
P3 noted that non-majors often face pressure to take CS courses due to fear of missing out. Majors had similar pressures, with P9 noting the self-imposed expectation that all majors take the accelerated introductory course, despite it being ``really hard if you haven't programmed before.'' P5 noticed how students often feel the pressure to ``pursue [the] selective or elite thing,'' namely, competitive pre-professional computing clubs.


\subsubsection{Institutional Culture} \label{sec:inst-culture} Institutional culture also risks negatively shaping \sob{}, with P2 and P8 describing downstream effects which inform student culture about what computing looks like. P8 also highlighted industry-centered institutional cultures, criticizing instructors who say ``You'll need this in the software industry,'' asserting that industry is ``all that matters.'' P3 discussed how a university's mission is core in establishing an academic culture, comparing academic (\ie{} retention and community building) and research (\ie{} knowledge production) missions. 

 



\subsubsection{Representation of Identity} \label{sec:representation}



Instructor representation can play a crucial role in belonging~\cite{rankin2020intersectional}. Participants viewed a lack of representation for both women and underrepresented minority (URM) groups, as well as stereotyped depictions of computer scientists in media (\eg{} the film \textit{The Social Network}, P4) as barriers to belonging (P1, P4, P6, P7, P9). P7 described how she felt empowered to ``[challenge the] perceptions of who does computer science'' while also not feeling pressure to singlehandedly break stereotypes. P6 noted how sufficient department-wide representation of female faculty can alleviate pressure to appear more feminine. P9 highlighted how greater female representation on course staff can support \sob{}. However, having identity representation alone is insufficient to build universal \sob{}: while P1 described her intention ``to appear approachable,'' she also noted how her ``visual identity markers'' by definition make her approachable to some students, but not all. 





\subsection{Student-oriented Interventions (RQ2)}
\label{sec:student-intervention}


\subsubsection{Individual Support}
Participants described approaches to student support, such as creating spaces for direct student engagement (P3, P4, P5, P7), holding 1:1 meetings (P2, P3, P4), and learning about student goals and personal interests (P2, P3, P4, P7). P2 mentioned the goal of these interventions: ``it's focusing less on why are you in the class, and more on what do you care about as a human being?'' Students also initiated conversations about their struggles with instructors they felt comfortable approaching (P3, P4, P5). When P1 reached out to struggling students, instead of asking for extensions, students appreciated the opportunity to collaboratively figure out next steps (\eg{} support needs, academic advising)---which surprisingly led P1 to provide fewer accommodations than expected, as students apologized and owned up to their behavior. 





\subsubsection{Course Engagement \& Interactions}
\label{sec:course-eng}
Participants also noticed decreased student engagement in recent years (P3, P4, P5, P6, P9). To increase engagement, participants offered extra credit to incentivize coming to class, interacting with peers, and attending office hours (P5, P6). TAs often led these spaces (\eg{} office hours, recitation section, labs), engaging with students more often than instructors due to scale (P3, P5, P7). Participants also had students form study groups based on shared interests (P5) and lab sections (P7).

\subsubsection{Community Spaces}
Community spaces facilitate course interactions, creating opportunities for students to ask questions and make friends (P3, P4, P5, P7). P4 described a department initiative for ``faculty hangouts,'' where students can meet with faculty, as well as his personal ``open office hours'' with ``LEGOs and snacks'' outside of regular course office hours. In addition, P4 described the value of these spaces: ``[students will] hear someone else's question, and then they'll go, `oh shoot, I actually had the same question.' And so I've, like, tricked them into learning.'' Both P3 and P4 noted awareness of service courses, student engagement programs, and broadening participation programs for First Generation Low Income students, where students can have a space to ``see other people like them.'' (P3) In contrast, P5 felt that such spaces focus too much time on imposter syndrome and less on technical content, arguing that incorporating content-specific support provides greater value. We found that, at both the individual- and course-level, participants who had more years of teaching experience employed a broader set of interventions.











\subsection{Course Design and Structure (RQ2)}
\label{sec:course-design}






\subsubsection{Content Relevance}
\label{sec:relevance}
P4 discussed how course content that appeals to student interests impacts belonging, finding it valuable to convince non-majors that computing affects their majors---for example, assignment integration based on political science, chemistry, and electrical engineering (P3, P4). P8 described his goal to be that ``every student [could find] at least one assignment that really resonates with them.'' Moreover, P2 described how he implemented collaborative course design with his students to create projects that were ``actually useful and beneficial for the world.'' Both P4 and P8 viewed their courses as not being a ``hard sell'' of CS, but rather that ``computing is kind of relevant'' to everyone. 

\subsubsection{Beyond Computing}
Several participants described how they perceive their roles to extend beyond teaching technical computing content. P8 described how he ``[tries] to normalize struggle [and] the fact that learning takes time,'' while P3 told struggling students ``[your performance] does not mean that you will not succeed in CS.'' Others described efforts to foster a growth mindset or lead by example by intentionally inserting bugs and debugging in class (P1, P4, P7). To help students succeed in college, participants taught ``softer skills or behavioral pattern strategies'' (P1) and focused on scaffolding implicit hidden curriculum topics (P6). 


\subsubsection{Course Staff}
\label{sec:course-staff}
In large courses, it can be hard for instructors to connect with all of their students. It is often TAs who address belonging~\cite{perlmutter-2023-field-accepted}, primarily due to their availability and approachability as peers. While hiring TAs, P6 explicitly told candidates, ``what's most important to me is you supporting a welcoming and inclusive classroom.'' P9 described how autonomy over course staff hiring can also be used to improve representation (as discussed in Section~\ref{sec:representation}) in the role models and peers students look up to. P3, P4, P5, and P6 mentioned the importance of building strong relationships with TAs, training them to take on the role of helping facilitate \sob{} in students (P4, P6). More specifically, P4 saw opportunities to utilize TAs from providing ``human feedback on everything'' to ``[laying] out community agreements and [give students] a chance to connect.'' 

\subsection{Barriers to Intervention (RQ3)}
\label{sec:barriers}




\subsubsection{Scale}
\label{sec:scale}
Scale was discussed by every participant as a significant barrier to belonging interventions. Whether class size (P3, P4, P5), course staff time and energy (P3, P4, P5), or physical space constraints (P7), scale impeded participants' ability to intervene in the ways that they would have liked to. P4 described how scale prevented accurate measurement of belonging, with P1 and P3 describing scale preventing further directed intervention efforts. Lastly, P8 saw how scale prevented student representation: ``I'm never gonna have everybody's identities and beliefs and perspectives reflected in [me or my TAs]...we're never going to hit everybody.''


\subsubsection{Student-TA Relationships}
\label{sec:student-ta}
Instructors have limited control over the environment and interactions between students and TAs. P3 and P8 discussed the necessity of TAs aligning with course messaging to foster belonging, with P8 noting that: ``[students will say,] `it's all great that [P8] said this, but here's six TAs who implicitly told me I don't belong here, so who am I supposed to listen to?''' Similarly, P7 described how TAs can act like ``jerks,'' due to a ``lack of empathy'' built up over time from students who demand answers. On the other hand, P6's TAs reported that students did not respect their non-binary pronouns, further straining TA-student relations.

\subsubsection{Societal Trends}
\label{sec:societal-trends}
P3 emphasized how ``the rise of AI has led to the deterioration of student learning communities and social interactions'' as students ``[turn] to AI because AI is non-judgmental [...] You don't have to go in person.'' P2 and P4 similarly acknowledged the joint impact of both AI and COVID-19 on social interaction and in-person engagement, due to both remote learning and reduced human support. Across institutions, participants noticed reduced enrollment---which P3 attributed to budget constraints---which was said to inhibit engagement (P1, P3, P5, P6, P9).



%% file: 05-new-discussion.tex
\section{Discussion}
\subsection{Comparing conceptualizations of belonging}
In addressing RQ1, we found that instructors' conceptualizations largely align with prior, student-centered conceptualizations of belonging~\cite{perlmutter-2023-field-accepted, mahar2013conceptualizing}, defining belonging through competence and self-efficacy, relatedness and community, and value and identity representations. Prior literature also emphasizes how structural inequities shape whether students feel that they belong in computing~\cite{rankin2021blacklong, rankin2020intersectional}. 

While instructors recognized the macro-scale influences on belonging, 
they conceptualized belonging through observable aspects of student experiences and concrete factors of educational environments in which they participated. Specifically, instructors discussed ``competence'' and ``self-efficacy'' through measurable metrics such as grades (Section~\ref{sec:prior-experience}), described ``relatedness'' and ``community'' in the context of institutional constructs such as distinctions between majors and non-majors (Section~\ref{sec:major}), explained `` value representation'' as the ideas and beliefs that students acquire from their peers and their department (Section~\ref{sec:student-culture}, ~\ref{sec:inst-culture}), and portrayed ``identity representation'' as how they appear and act in front of students (Section~\ref{sec:representation}). Similarly, instructors' interventions to support \sob{} were grounded in what they can directly change (Section~\ref{sec:student-intervention}, ~\ref{sec:course-design}).

This comparison suggests that while recognizing the importance of intersectional approaches to addressing inequities in belonging is critical~\cite{runa2025sense}, instructors will also need to understand how to translate abstract principles into actionable practices within their spheres of influence to realize the intended impact of such approaches. Therefore, there should be greater efforts in helping instructors connect observed course-level patterns to the deeper structural dynamics that produce inequitable belonging in computing education. One such practice is reflexivity, which has been shown to surface productive insights into how individuals with power in the classroom can act upon structural inequities, specifically by interrogating how their own assumptions, course and instructional norms may reproduce barriers to belonging~\cite{brooksramirez2026pluriversal}.

\subsection{Mobilization of TAs in supporting belonging}
One notable finding for RQ2 and RQ3 is how instructors' described scale as a barrier to supporting belonging (Section~\ref{sec:scale}), resulting in mobilizing TAs as intermediaries between them and students (Section~\ref{sec:course-staff}). This is consistent with prior work showing that large computing courses increasingly depend on TAs to provide individualized support, staff office hours and recitation section, and maintain course infrastructure~\cite{luo-student-support-2026, perlmutter-2023-field-accepted, alvarado2017microclasses, kanchi-systems-2026}. Instructors noted TAs as central to reaching students and facilitating community spaces, especially with P1 describing difficulties in student approachability (Section~\ref{sec:course-eng},~\ref{sec:course-staff})~\cite{perlmutter-2023-field-accepted}.


However, instructors' reliance upon TAs also introduced deep tensions. Instructors identified instances where TAs were inconsistent in course messaging, told students implicitly that they did not belong in computing, and lacked empathy to support students' needs (Section~\ref{sec:student-ta}). While TAs can expand instructors' capacity to support belonging at scale, they can also oppose these aspirations when their roles are not carefully supported or aligned.

These findings suggest that TA training in large computing courses should explicitly consider students' \sob{}, rather than viewing it as a side effect of supporting students. While TA training is common in practice~\cite{yan2025teaching, kanchi-systems-2026}, instructors emphasized the need for TA training that prepares TAs to facilitate belonging (Section~\ref{sec:course-staff}). Such training should include aligned course messaging, shared learning goals, and strategies for responding to students' academic and social needs~\cite{brooksramirez2026pluriversal, yan2025teaching}. Belonging interventions must not only focus on faculty actions, but actions of the instructional staff, through whom students most often engage with the course.

\subsection{Balancing student expectations}

When answering RQ2, we found that instructors perceived students' expectations as an influential factor in course design and belonging interventions. These expectations concerned not only what students expected to learn, but also whether they saw computing as personally relevant, approachable, and achievable. Instructors therefore discussed a delicate balance between responding to students' expectations and advancing course goals.




Instructors saw students' majors influence expectations about computing relevancy, which they described as crucial to cultivating belonging (Section~\ref{sec:major},~\ref{sec:relevance}). For example, P3 and P4 designed assignments around students' disciplines, consistent with prior work showing that solving discipline-specific problems can help non-majors better understand the reach of computing~\cite{dawson-designing-2018}. However, instructors also noted that calibrating relevance across students with different majors, goals, and experience was difficult at scale (Section~\ref{sec:scale}). Furthermore, P7, who taught non-majors, noted that many students had no aspirations of belonging in computing.

Students also were said to have expectations about who can do computing. These expectations were reinforced by stereotypes (P4)~\cite{lewis2016dontcode} or limited representation of students' identities and values~\cite{lewis-alignment-2019}. Instructors noted difficulties in representing and being approachable to every student---creating a gap between students' definitions of belonging and the forms of belonging instructors could realistically support (Section~\ref{sec:representation}). This gap is especially consequential for marginalized students, who may already encounter narrow representations of who a computer scientist is~\cite{lewis2016dontcode}.

Finally, instructors described students' expectations of success as shaped by external narratives and cultures. Institutional pathways could discourage non-majors from studying CS (Section~\ref{sec:major}), echoing prior work showing that competitive enrollment policies can negatively impact belonging for students without prior experience~\cite{nguyen-2020-competitive}. Both institutional and student cultures were said to be influenced by industry narratives that portrayed computing as selective, elite, or narrowly tied to software engineering, resulting in students enrolling in accelerated courses or applying for competitive pre-professional computing clubs (Section~\ref{sec:inst-culture},~\ref{sec:student-culture}). When students' expectations are shaped by normative career practice, they may struggle to see alternative forms of participation,
leaving less space to pursue their aspirations~\cite{Kirdani_Ryan_2023}.




These findings suggest that expectation-setting can function as a belonging intervention, shaping how students perceive computing and who can be successful in the field. 
By helping students see computing as relevant, approachable, and achievable, instructors can support broader forms of belonging that account for students' intersecting institutional, disciplinary, and personal contexts.






\subsection{Larger societal trends introduce new tensions in belonging}
In response to RQ3, we found instructors describing students' \sob{} to be increasingly shaped by uncontrollable societal forces. In particular, instructors noted contemporary shifts in support-seeking and participation associated with generative AI and the lingering effects of the COVID-19 pandemic (Section~\ref{sec:societal-trends}). Several described students as turning to AI for accessible and non-judgmental support (P3), rather than seeking course staff~\cite{perlmutter-2023-field-accepted}. Instructors connected this shift to reduced in-person participation and fewer opportunities to engage with classmates and course staff~\cite{hou2025all}, which has shown to reduce students' belonging~\cite{stebleton2014recent, markel2020designing, mooney-2020-belonging, runa2025sense}.
However, remote learning did not affect all students uniformly; for some groups, including minoritized women, remote environments could feel safer and more inclusive, thereby increasing belonging~\cite{mooney-2020-belonging, runa-2023-student-gender}. 

These trends reduced instructors' visibility into students' experiences, increasing difficulty in interpreting student needs, measuring \sob{}, and identifying when and how to intervene (Section~\ref{sec:measuring}). In response, instructors organized community spaces to increase engagement and connection, and encouraged more 1:1 interactions to connect with students on a deeper level (Section~\ref{sec:student-intervention}). 


These findings suggest a tension: modern technologies and instructional modalities can make courses more accessible and inclusive, while also reducing human interaction through which belonging is often cultivated~\cite{mooney-2020-belonging, runa2025sense}. Belonging and inclusivity are not opposites, but they require different design considerations under changing societal and technological conditions. Instructors must therefore foster meaningful course interactions that increase the visibility of students' needs while adapting to these broader shifts. However, such investment can intensify instructor burden. One opportunity is for instructors to initiate community opportunities while building capacity for students and TAs to help sustain them, such as TA-led study sessions or allocating funds for student-oriented activities~\cite{yan2025teaching, rankin2021blacklong}.



%% file: 06-conclusion.tex
\section{Limitations}
Our small convenience sample ($n=9$) pool came from four large U.S. public universities. Additional interviews may have surfaced new or different themes. Lecture-based, TA-supported courses in our sample may differ from smaller classes, upper-division courses, private institutions, and other regions. Voluntary participation may favor instructors attentive to belonging, while recruitment through researchers’ former institutions may introduce selection bias. These limitations may reduce the generalizability of our findings. 

\section{Conclusion \& Future Work} 
We conducted semi-structured interviews with nine introductory computing instructors' and identified tensions in aligning prior work with instructor perceptions, mobilizing TAs, balancing student expectations, and adapting to societal shifts. We encourage computing instructors to design belonging interventions that consider these tensions and the varied contexts in which students experience belonging. Future work can extend our work by conducting larger studies that engage instructor perceptions of student belonging alongside the broader contexts of students and TAs.